\documentstyle[epsf,12pt]{article}
\title{Soft dipole mode in $^{11}$Li and three body continuum}
\author{Yu. A. Lurie and {\underline{ A. M. Shirokov}}\thanks{Talk presented
at the Conference "Halo-94" (Copenhagen, January 1994)}\\
Institute for Nuclear Physics at Moscow State University, \\
Moscow 119899, Russia \\
and \\
Yu. F. Smirnov\thanks{On leave of absence from Institute for Nuclear
Physics at Moscow State University, Moscow 119899, Russia.}\\
Instituto de F\'{\i}sica, UNAM, Apartado postal 20-364,\\
Delegation Alvaro Obregon, 01000 M\'{e}xico D.F., M\'{e}xico}
\date{ }
\begin{document}
\maketitle
\setcounter{footnote}{2}
\begin{abstract}
Properties of the neutron rich $^{11}$Li nucleus are calculated in
the framework of the cluster model $^{9}$Li $+n+n$. The formalism of the
harmonic
oscillator representation is used for the description of bound and continuum
spectrum states in the three-body-democratic-decay approximation. It is shown
that this approach allows one  to take into account
adequately the long asymptotic tail of the $^{11}$Li wave function
({\em neutron
halo}) and to reproduce correctly the binding energy, radius and $^{11}$Li
electromagnetic dissociation cross-section on target nuclei. The shape and
the energy
position of the ${\cal B}(E1)$ peak corresponding to the soft dipole mode
are also in agreement with experiment.
\end{abstract}
\section*{Introduction}
\par
Recently secondary beams of radioactive heavy ions become
available. As a result one get a tool for experimental studies of light
neutron-excess $\beta$-unstable nuclei. $^{11}$Li is one of the
most interesting nuclei of the type, and its properties have
being intensively studied both experimentally \cite{1}--\cite{cc4}  and
theoretically (see, e.g., \cite{c17}--\cite{c15} and references therein).
\par
The main peculiarity of the $^{11}$Li nucleus is
the so-called {\em neutron halo} formed by two weakly-bound
neutrons (two-neutron separation energy $\varepsilon_{2n}=247\pm80$ keV;
note, that there are no bound states in two-body subsystems
($^{9}$Li$+n$) and ($n+n$) \cite{c16}). As the binding energy
of the neutron pair is small,  the wave function decreases slowly at large
distances. It results in the anomalously large value of the $^{11}$Li
r.m.s.\ radius,
${\bf <}r^{2}{\bf >}^{1/2}_{11}=3.16\pm0.11$ fm \cite{3} (to be
compared to the one of $^{9}$Li,
$\ {\bf <}r^{2}{\bf >}^{1/2}_{9}=2.32\pm0.02$ fm \cite{c16}). Evidently, the
anomalously large electromagnetic dissociation (ED) cross
section of $^{11}$Li beam on heavy target nuclei is the
manifestation of the neutron halo (the $^{11}$Li ED cross section on the
Pb target is 20 times the one of the carbon beam \cite{4}). To explain this
effect it was suggested \cite{c17,c18} that oscillations of the halo
neutrons with respect to the core $^{9}$Li give rise to the specific
low energy branch of the giant dipole resonance (soft dipole
mode) with excitation energy less then 1 MeV. Soft dipole mode
is supposed to exhaust about 10\% of the dipole energy-weighted
sum rule (EWSR).
\par
The shell-model \cite{6,7,11,15,lenske}, cluster
\cite{9,13,14,lurie,Danilin,c15}
and combined
cluster-shell-model \cite{8,10,c15} approaches have been used in theoretical
studies of the $^{11}$Li neutron halo properties. In some  of these
calculations (see, e.g., \cite{6}--\cite{10} a variational approach
has been used which does not allow for
continuum spectrum effects. There are also calculations with an
allowance for two- \cite{5,11,lenske} or
three-body \cite{13,14,lurie,Danilin,c15} continuum.
\par
 In this contribution we present the study of the
neutron halo properties of
$^{11}$Li in a three-body cluster model $^{9}$Li$+n+n$.
We make use of a three-body wave function expansion in six-dimensional
harmonic oscillator eigenfunctions. Our attention is focused on the
effects of three-body continuum. The continuum spectra effects
are accounted for in the framework of the oscillator
representation of the scattering theory \cite{16}. This method has
been generalized \cite{17} on the case of the true three-body scattering
\cite{18,19}
and has been successfully used recently in the study of the $^{12}$C monopole
excitations in the cluster model $\alpha+\alpha+\alpha$ \cite{20}.
As it has been
noted, there are no bound two-body subsystems like ($^{9}$Li$+n$)
or ($n+n$) in the
system. Therefore, only the so-called "democratic" three-body decay
\cite{19,21} is
inherent to this system. It can be adequately described in the framework of
the "true" three-body scattering theory (democratic decay approximation)
\cite{19}. In the study of the
ground and
excited states properties we go beyond the pure diagonalization of the
Hamiltonian matrix on the basis of oscillator functions. We
search for the $S$-matrix pole corresponding to the ground state,
and construct an infinite series expansion in oscillator
functions for the ground and excited state wave functions. It
enables us not only to calculate the $^{11}$Li ground state energy
with high accuracy, but to describe also the wide spatial
distribution of two valence neutrons in $^{11}$Li (neutron halo).\\
\section{The model}
\par
The $^{11}$Li ground state and the continuum spectrum wave
functions are calculated in the framework of the three-body
cluster model $^{9}$Li$+n+n$. The model assumptions are the following ones:
\par
1) The cluster $^{9}$Li is supposed to be structureless and the
excitations of its internal degrees of freedom are not
considered.
\par
2) We don't account for non-central components of the
interaction between two valence neutrons and between valence
neutron and the cluster $^{9}$Li. Therefore, the wave function can
be characterized by the three-body
orbital angular momentum $J$ and its projection $M$.
\par
3) The states with the total spin of the valence neutron pair
$S=0$ are only considered, and the ground state three-body orbital
angular momentum is supposed to be equal to zero: $L=0$.
\par
4) $n$-$^9$Li interaction is described by the shallow
potential of Johansen et al \cite{9} effectively simulating the
Pauli principle \cite{Jens}. $NN$-interaction is described by the
Gaussian potential~\cite{9}.
\par
5) Only democratic decay channels are allowed for.
\vspace{4mm}
\par
The wave function of the system $^{9}$Li$+n+n$,
$\psi_{JM}({\mbox {\bf x}, {\bf y}}) $, is expanded in three-body
hyperspherical functions,
$\Phi^{l_{x}\,l_{y}\,J\,M}(\hat{\rho})$,
\begin{eqnarray}
\psi_{JM}({\mbox {\bf x}, {\bf y}}) = \sum_{K\,l_{x}\,l_{y}}
\psi^{(J)}_{K\,l_{x}\,l_{y}}(\rho) \,
\Phi^{l_{x}\,l_{y}\,J\,M}_{K}(\hat{\rho})
\, ,    \label{1}
\end{eqnarray}
where $K$ is
hypermomentum, $l_{x}$ and $l_{y}$ are the angular momenta
corresponding to the Jacobi coordinates
\begin{eqnarray}
{\mbox {\bf x}} = \sqrt{ \frac {m\omega}{2\hbar} } ({\mbox {\bf r}}_{1}\;-\;
{\mbox {\bf r}}_{2})\,,
\hspace{2em}{\mbox {\bf y}} = \sqrt{ \frac{18}{11}\frac {m\omega}{\hbar} }
(\frac {{\mbox {\bf r}}_{1}+{\mbox {\bf r}}_{2}}{2}\;-\;
{\mbox {\bf r}}_{3})\,,
 \label{2}
\end{eqnarray}
respectively, $m$ is the neutron mass, $\mbox{\bf r}_{i}$ are
coordinates of the valence neutrons ($i=1,2$) and the cluster
$^{9}$Li ($i=3$), {\bf $\rho$}$=({\mbox {\bf x}}^{2}+{\mbox {\bf
y}}^{2})^{1/2}$ is a three-body hyperradius, $\hat{\rho}$ stands
for the set of angular variables in the six--dimensional space.
\par
In the c. m. frame the Hamiltonian is of the form:
\begin{eqnarray}
H = T + V_{12} + V_{13} + V_{23}\;,        \label{3}
\end{eqnarray}
where $T$ is the three-body relative motion kinetic energy operator,
and $V_{ij}$ are the two-body potentials. For the radial wave
functions $\psi^{(J)}_{K\,l_{x}\,l_{y}}(\rho)$ we have the usual set of the
$K$-harmonic method coupled equations (see, e.g., \cite{19}).
The equations are solved by expanding the radial wave function,
\begin{eqnarray}
\psi^{(J)}_{K\,l_{x}\,l_{y}}(\rho) = \sum_{n=0}^{\infty} \;
D^{(J)}_{n\,K\,l_{x}\,l_{y}}(E) \ \varphi_{n\,K}(\rho)\,,    \label{4}
\end{eqnarray}
in the six-dimensional harmonic oscillator eigenfunctions,
\begin{eqnarray}
\varphi_{n\,K}(\rho) = (-1)^{n} \sqrt{ \frac{2\,n!}{\Gamma(n+K+3)} } \;
\rho^{K} \; L^{K+2}_{n}(\rho^{2}) \; \exp(-\rho^{2}/2) \;,      \label{5}
\end{eqnarray}
where $L^{\alpha}_{n}(x)$ is the associated Laguerre polynomial.
\par
For large values of the total number of oscillator quanta,
$N=2n+K$, the potential energy matrix elements are small compared
to the matrix elements of kinetic energy which increase
linearly with $N$. In the framework of the oscillator
representation of scattering theory the matrix of the potential energy,
$V=V_{12}+V_{13}+V_{23}$, on the basis of functions {\mbox (\ref{5})}
is truncated, i.e. all matrix elements
$V^{N'\Gamma'}_{N\Gamma}$ for $N>\tilde{N}$ and/or $N'>\tilde{N}$
are neglected ($\tilde{N}$ denotes the truncation boundary).
The kinetic energy matrix, $T^{N'\Gamma'}_{N\Gamma}$,
is an infinite tridiagonal matrix for any value of the multi-index
$\Gamma=\{K,l_{x},l_{y}\}$ which labels different decay channels
in the hyperspherical representation. The eigenvectors of this
matrix which are important for the description of the wave
function asymptotical behavior at $\rho \longrightarrow \infty$,
 in the oscillator representation are of the form \cite{17}:
\begin{eqnarray}
& D^{(J)\;(\Gamma')}_{n\,K\,l_{x}\,l_{y}}(E) =
\frac{1}{2} [ \; \delta_{\Gamma\,\Gamma'} \;
C^{(+)}_{n\,K}(E) \ -\ S_{\Gamma\,\Gamma'}\;
C^{(-)}_{n\,K}(E) \;] \ \ \ \ & for\ E\geq0,  \\   \label{6}
& D^{(J)}_{n\,K\,l_{x}\,l_{y}}(E) = \alpha_{\Gamma}\;
C^{(+)}_{n\,K}(E) \ \ \ \
& for\ E<0,     \label{7}
\end{eqnarray}
where $S_{\Gamma\,\Gamma'}$ is the matrix element of the $S$-matrix,
and $\alpha_{\Gamma}$ is the normalization constant of the bound state
wave function for the channel $\Gamma$.
\par
There are analytical expressions for the functions $C^{(\pm)}_{n\,K}(E)$
\cite{17},
\begin{eqnarray}
C^{(\pm)}_{n\,K}(E) = C_{n\,K}(E) \; \pm\; i\,S_{n\,K}(E)\,,       \label{8}
\end{eqnarray}
where
\begin{eqnarray}
& &S_{n\,K}(E) = \frac{1}{\sqrt{\hbar\omega}} \;
\sqrt{\frac{2\,n!}{\Gamma(n+K+3)}} \; q^{K+2} \; L^{K+2}_{n}(q^{2}) \;
\exp{(-\frac{q^{2}}{2})}\;, \\     \label{9}
& &C_{n\,K}(E) = -\frac{1}{\pi\,S_{0\,K}(E)} \ P.V.\int_{0}^{\infty}\;dE'\,
\frac {S_{n\,K}(E')\,S_{0\,K}(E')} {E\ -\ E'} \;,  \\     \label{10}
\end{eqnarray}
and the total energy of the system, $
E = \frac{1}{2}q^{2} \hbar\omega
$.
The functions $S_{n\,K}(E)$ and $C_{n\,K}(E)$ give rise to functions with the
following asymptotics in the coordinate space:
\begin{eqnarray}
\sum_{n=0}^{\infty} \; S_{n\,K}(E) \ \varphi_{n\,K}(\rho) =
 \frac {1}{ \sqrt{\hbar\omega} \rho^{2} } \; J_{K+2}(q\rho )\,, \\
 \label{12}
\sum_{n=0}^{\infty} \; C_{n\,K}(E) \ \varphi_{n\,K}(\rho) \simeq
\frac {1}{ \sqrt{\hbar\omega} \rho^{2} } \; N_{K+2}(q\rho )\,,
\label{13}
\end{eqnarray}
\par
The asymptotic expressions {\mbox (\ref{6})} for
$D^{(J)\; (Gamma')}_{n\,K\,l_{x}\,l_{y}}(E)$
correspond to the account of democratic
decay channels only \cite{19}, i.e. the wave function in the asymptotic
region
is a superposition of an outgoing six-dimensional spherical wave in
the channel $\Gamma'$ and ingoing six-dimensional spherical waves in
all channels $\Gamma$. The applicability of the democratic decay
approximation for various nuclear reactions with few particles
in a final state has been discussed in \cite{19,21}. In the
framework of the oscillator representation of scattering theory
this approximation has been used \cite{20} in the study of monopole
excitations of $^{12}$C in the cluster model $\alpha+\alpha+\alpha$.
The analysis of experimental data for $2^{+}$ states in $A=6$ nuclei within
the $\alpha+N+N$ cluster model assumptions has been performed using this
approximation in ref. \cite{22}. This analysis has demonstrated that the
approximation is an adequate tool for description of the decay states of
Barromean nuclei.
\par
The energy spectrum and the wave functions are calculated
in the following way \cite{17}. First of all, one should find
eigenvalues, $E^{(J)}_{\lambda}$, and eigenvectors,
$\{\gamma^{(\lambda)\;(J)}_{n\,\Gamma}\}$, of the truncated
Hamiltonian matrix, $\{H^{n'\Gamma'\;(J)}_{n\,\Gamma}\}$,
$2n+K\leq\tilde{N}$,
$2n'+K'\leq\tilde{N}$; i.e. one should solve an eigenproblem
for the set of equations
\begin{eqnarray}
\sum_{n',\ \Gamma'}^{ 2n'+K'\leq\tilde{N} }
\  [ H^{n'\,\Gamma'\;(J)}_{n\,\Gamma} \ -\
\delta_{n\,n'}\;\delta_{\Gamma\,\Gamma'}\;
E^{(J)}_{\lambda}]\ \gamma^{(\lambda)\;(J)}_{n'\,\Gamma'}
\ =\ 0\;,\ \ \  2n+K\leq\tilde{N}\,.       \label{14}
\end{eqnarray}
\par
In the usual variational approach, the lowest eigenvalue,
 $E^{(J=0)}_{0}$, and
the corresponding eigenvector,
$\{\gamma^{(0)\;(J=0)}_{n\,\Gamma}\}$ are treated as the ground state energy
and the corresponding wave function in the oscillator representation. Such
approximation is unable to describe the slowly dying asymptotic
tail of the halo neutron space distribution. Thus, in order to
describe the neutron halo properties of the $^{11}$Li nucleus, one
is pushed to account for the asymptotic region $N\geq\tilde{N}$. For bound
states this account is equivalent to the location of $S$-matrix
poles. The $S$-matrix can be calculated by the equation \cite{17}
\begin{eqnarray}
S = (A^{(+)})^{-1} A^{(-)}\; .   \label{15}
\end{eqnarray}
The matrices $A^{(+)}$ and $A^{(-)}$ have the following matrix elements:
\begin{eqnarray}
A^{(\pm)}_{\Gamma\,\Gamma'} =
\delta_{\Gamma\,\Gamma'}\;C^{(\pm)}_{\tilde{n},\,K}(E)
\ + \ P^{\tilde{n}',\,\Gamma'}_{\tilde{n},\,\Gamma}(E)\;
T^{\Gamma'}_{\tilde{n}',\,\tilde{n}'+1}\;C^{(\pm)}_{\tilde{n}'+1,\,K'}(E)\;,
\label{16}
\end{eqnarray}
the kinetic energy matrix elements,
$T^{\Gamma}_{n,\,n+1} = -\frac{1}{2}\hbar\omega \;
\sqrt{(n+1)(n+K+3)}$, $\tilde{n}=\frac{1}{2} (\tilde{N}-K)$,
$\tilde{n}'=\frac{1}{2}(\tilde{N}-K')$,
and
\begin{eqnarray}
P^{n',\,\Gamma'}_{n,\,\Gamma}(E) =
\sum_{\lambda}\frac {\gamma^{(\lambda)}_{n'\,\Gamma'} \;
\gamma^{(\lambda)}_{n\,\Gamma} } { E_{\lambda}\;-\;E } \;,     \label{17}
\end{eqnarray}
\par
To calculate the bound state energy, i.e. to locate the corresponding
$S$-matrix pole, one should solve the nonlinear equation \cite{17}
\begin{eqnarray}
\det A^{(+)} = 0\,,     \label{18}
\end{eqnarray}
which can be easily obtained from {\mbox (\ref{15})}.
Asymptotic normalization constants of  the bound states, $\alpha_{\Gamma}$,
can be found
by numerical solution of the set of linear homogeneous equations
\begin{eqnarray}
\sum_{\Gamma'}\,A^{(+)}_{\Gamma \Gamma'}\;\alpha_{\Gamma'} = 0 \; .
\label{20}
\end{eqnarray}
The set of $\alpha_{\Gamma}$ can be multiplied by any common multiplier.
This uncertainty
is eliminated by numerical normalization of the wave function. Now
the coefficients $D^{(J)}_{n\,K\,l_{x}\,l_{y}}(E)$ of the expansion
{\mbox (\ref{4})} in the asymptotic region $N\geq\tilde{N}$
can be easily obtained using {\mbox (\ref{7})}.
\par
For the continuum spectrum states we calculate $S$-matrix for any
positive energy $E$ using
(\ref{15})--(\ref{17}), and than calculate the set of
$D^{(J)}_{n\,K\,l_{x}\,l_{y}}(E)$  in the asymptotic region
$N\geq\tilde{N}$
using {\mbox (\ref{6})}.
\par
The coefficients $D^{(J)}_{n\,K\,l_{x}\,l{y}}(E)$ in the inner region
$N\leq\tilde{N}$ are calculated by the expression:
\begin{eqnarray}
D^{(J)}_{n\,K\,l_{x}\,l_{y}}(E) =
- \sum_{\Gamma'}\;P^{n\,\Gamma}_{\tilde{n}'\,\Gamma'}(E)\;
T^{\Gamma'}_{\tilde{n}',\,\tilde{n}'+1}\;D^{(J)}_{\tilde{n}'+
1,\,K'\,l'_{x}\,l'_{y}}(E)\;,
\label{19}
\end{eqnarray}
where $\tilde{n}'=\frac{1}{2}(\tilde{N}-K')$ and for calculation of
$D^{(J)}_{\tilde{n}+1,\,K\,l_{x}\,l_{y}}(E)$
eqs. (\ref{6}) or (\ref{7}) are used for continuum or bound state,
respectively.\\
\section{The results of the calculations}
\par
The interactions of the valence neutrons with each other and
with the cluster $^{9}$Li are described by the potentials $V_{12}(r_{12})$
and $V_{13}(r_{13})=V_{23}(r_{23})$, respectively. We use the following
parametrization of the potentials \cite{9}:
\begin{eqnarray*}
&V_{ij}(r) =  V^{(1)}_{ij}\,\exp{[-(r/b^{(1)}_{ij})^{2}]}\ +
\ V^{(2)}_{ij}\,\exp{[-(r/b^{(1)}_{ij})^{2}]}\,, \\
&V^{(1)}_{12} = -31\ \mbox{MeV,}\hspace{3em} V^{(2)}_{12} =  0,
\hspace{3em} b^{(1)}_{12} = 1.8\ \mbox{fm;} \\
&V^{(1)}_{13} =  -7\ \mbox{MeV,}\hspace{2em} V^{(2)}_{13} = -1\ \mbox{MeV,}
\hspace{2em} b^{(1)}_{13} = 2.4\ \mbox{fm,} \hspace{2em} b^{(2)}_{13}
= 3.0\ \mbox{fm.}
\end{eqnarray*}
\par
In the external asymptotic region $N\geq\tilde{N}$ we consequently
allow for channels $\Gamma $ characterized by
$K=K_{min},\,K_{min}+2,\ldots$ ($K_{min}$
is the minimal possible value of $K$ for a given $J$) until the
convergence for all physical properties under consideration is
achieved. The convergence is found to be very good, and the
allowance for the decay channels with $K>K_{min}+2$ do not yield any
visual variation of the results. So, we consider in the external
asymptotic region $N>\tilde{N}$ the channels with $K\leq K_{min}+2$ only.
Note,
that components with all possible values of $K\leq \tilde{N}$ are accounted
for in the calculation of the wave function in the inner region
$N\leq \tilde{N}$.
\par
The parameter $\hbar\omega$ is set to be equal to 7.1 MeV in our calculations.
This value corresponds approximately to the minimum of ground state energy
$E_{0}$.\\
\subsection{The ground state}
\par
The results for the $^{11}$Li ground state for different values
of the truncation parameter $\tilde{N}$ are presented in the table 1.
The variational
ground state energies,  $E^{(d)}_{0}$, obtained by the pure
diagonalization of the truncated Hamiltonian matrix are listed
in the second column, while the $J$-matrix results, $E_{0}$, which are the
solutions of the eq. \mbox{(\ref{18})} are listed in the third column.
It is seen, that by locating the $S$-matrix pole using eq. \mbox{(\ref{18})}
that is equivalent to the allowance for the long
asymptotic tail of the wave function, we improve essentially the
convergence for the binding energy.
\begin{table}[thb]
\centering
\caption{$^{11}$Li ground state properties (see text for details).}
\begin{tabular}{|c|c|c|c|c|} \hline \hline
                     &     \multicolumn{2}{c|}{Ground state energy,}      &
\multicolumn{2}{c|}{Neutron halo}                      \\
 Truncation          &     \multicolumn{2}{c|}{MeV}                       &
\multicolumn{2}{c|}{mean square radius}                \\
 boundary $\tilde{N}$&     \multicolumn{2}{c|}{}                          &
\multicolumn{2}{c|}{$<r^{2}>^{1/2}_{11}$, fm}          \\ \cline{2-5}
                     &       $E_{0}^{(d)}$        &        $E_{0}$        &
$<r^{2}>^{1/2 \ (d)}_{11}$   &   $<r^{2}>^{1/2}_{11}$   \\ \hline
   12                &          -0.012            &         -0.150        &
            2.83            &         3.31             \\
   16                &          -0.116            &         -0.199        &
            2.91            &         3.29             \\
   20                &          -0.171            &         -0.225        &
            2.98            &         3.31             \\
   24                &          -0.202            &         -0.240        &
            3.04            &         3.32             \\ \hline
 Experiment          &   \multicolumn{2}{c|}{-0.247$\pm$0.080}            &
 \multicolumn{2}{c|}{3.16$\pm$0.11}                    \\ \hline \hline
\end{tabular}
\\
\caption{The effect of the Lanczos smoothing on the ground state energy of
$^{11}$Li (see text for details).}
\begin{tabular}{|c|c|c|c|c|c|c|c|c|} \hline \hline
$\tilde{N}$     & 10    & 12    & 14    & 16    & 18    & 20    & 22    & 24
\\ \hline
$E_0$           &-0.119 &-0.150 &-0.181 &-0.199 &-0.215 &-0.225 &
&-0.240 \\ \hline
$E_{0}^{(w.s.)}$&-0.136 &-0.266 &-0.204 &-0.262 &-0.232 &-0.262 &-0.249
&-0.263 \\ \hline \hline
\end{tabular}
\\
\end{table}
\par
The results presented in the table 1 have been obtained using Lanczos
smoothing of the
three-body potential energy matrix \cite{hung}. It has been shown in ref.
\cite{hung} that the
smoothing improves the convergency of calculations of two-body
scattering phases.
 The results of the full $J$-matrix calculations of the ground
state energy of $^{11}$Li obtained with the use of the smoothing, $E_0$,
and without it,
$E_{0}^{(w.s.)}$, for various values of the truncation boundary
$\tilde{N}$ are presented in
table 2. It is seen from the table 2 that the smoothing causes
underestimation of the binding
energy of a three-body system. Nevertheless,  $E_{0}^{(w.s.)}$
shows a staggering as
$\tilde{N}$ increases\footnote{Note, that $J$-matrix calculation is
not a variational one, so,
the ground state energy may behave non-monotonically as
$\tilde{N}$ increases.}. The
smoothing results in a usual-type monotone decrease of $E_0$ as
$\tilde{N}$ increases
and in a better convergence of continuum spectrum calculations.
At the same time,
the ground state wave functions obtained with the smoothing and without
it differ very
slightly. So, below we shall discuss only the results of calculations
with the smoothed
potential energy matrix. The role of the smoothing in three-body
calculations we shall
discuss in more detail elsewhere \cite{lurie}.
\par
The $^{11}$Li r.m.s.\ radius, $<r^{2}>^{1/2}_{11}$, can be calculated by
the following equation:
\begin{eqnarray}
<r^{2}>_{11} = - \frac{9}{11} <r^{2}>_{9} +
\frac{\hbar}{11\,m\,\omega} <\rho ^{2}> \,,     \label{21}
\end{eqnarray}
where $<r^{2}>^{1/2}_{9}$ is the $^9$Li r.m.s.\ radius
and the mean square value of the hyperradius, $<\rho ^{2}>$,
can be easily calculated using  the ground state wave function.
The values of $<r^{2}>^{1/2 \ (d)}$ and
$<r^{2}>^{1/2}$ obtained by the
pure diagonalization of the truncated Hamiltonian matrix and
with the allowance for the asymptotic tail of the wave function,
respectively, are presented in the 4-th and the 5-th columns of the table~1.
\par
Figure 1 presents the transverse momentum distribution of
the cluster $^{9}$Li in $^{11}$Li. This momentum distribution is
currently supposed (see, e.g., \cite{12}) to be proportional to the
$^{9}$Li transverse momentum distribution $\frac{dN}{dp_{\bot}}$ in the
fragmentation of high-energy $^{11}$Li beams on target nuclei. The
experimental data for $^{11}$Li fragmentation \cite{2} are also presented
on the figure. The results of $J$-matrix calculations with $\tilde{N}=16$
and $\tilde{N}=24$ are so close to each other that their plots on fig.1
coalesce and cannot be distinguished.
\begin{figure}[bth]
\epsfverbosetrue
\epsfbox{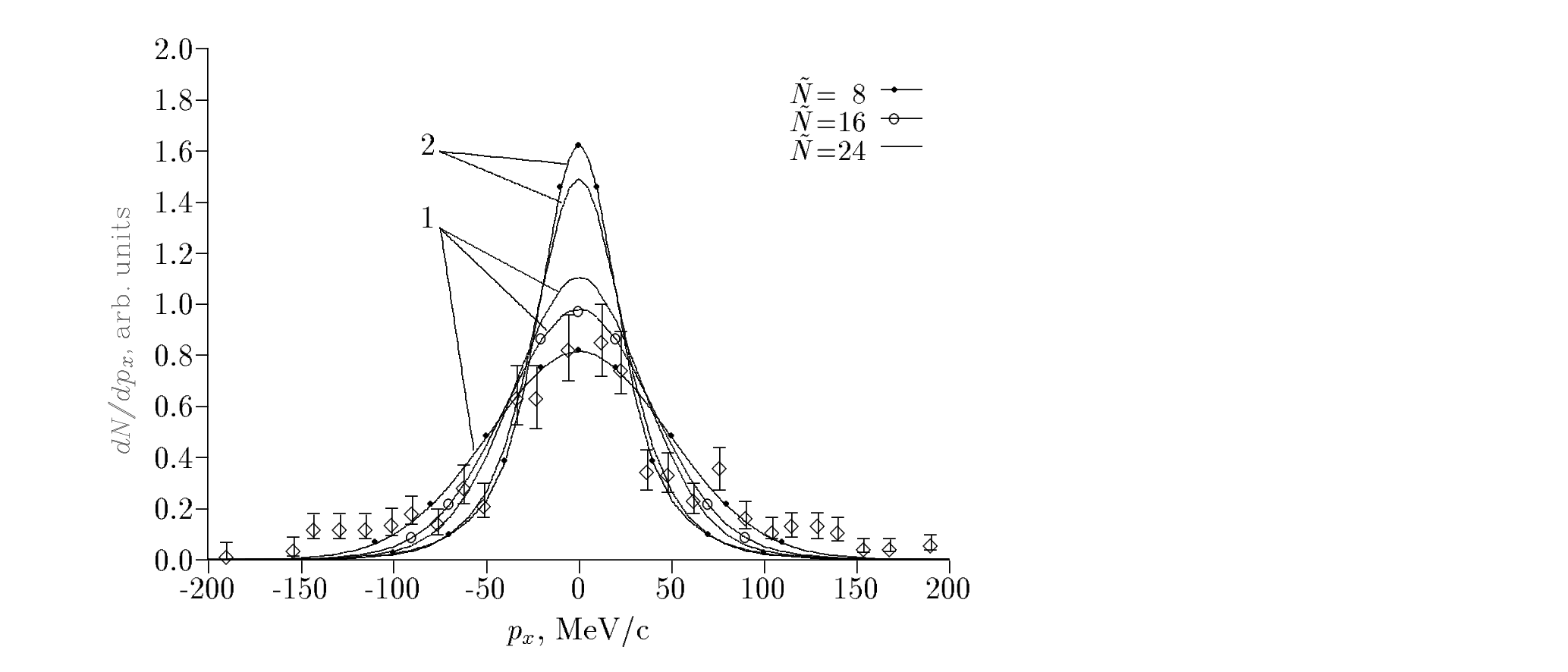}
\caption{The $^{9}$Li transverse momentum distribution in the ground state
of $^{11}$Li. 1 --- variational calculations, 2 --- $J$-matrix calculations.
Experimental data for the $^{9}$Li transverse momentum
distribution in the fragmentation of $^{11}$Li 800 MeV/nucleon beam are
taken from ref.\ \protect\cite{2}.}
\end{figure}
\par
It is seen that in calculation of the ground state, the allowance for the
wave function asymptotics is very important for a weekly-bound system like
$^{11}$Li. The terms of expansion (\ref{4}) with the
number of total oscillator quanta $N\simeq 100$ that cannot be obtained
in the usual oscillator-basis variational calculations, play an
essential role in the formation of the transverse momentum distribution,
r.m.s.\ radius, etc.
The convergence of $<r^{2}>^{1/2}$, transverse momentum
distribution and other properties of the wave function (e.g., of the weights
of its components)  in the full $J$-matrix
calculation is very good. Nevertheless, it is seen that
the r.m.s.\ radius
converges to a value that is somewhat larger than the experimental
one, and the calculated transverse momentum distribution
is narrower than the experimental one. These shortcomings can
be  overcome by the adjustment of $n$--$^{9}$Li potential. We
have not aimed to fit the potential to the $^{11}$Li
properties, we have just take its parameters from ref.\ \cite{9}. The
differences between the calculated and experimental values of
r.m.s.\ radius arise from the
fact that in ref.\ \cite{9} the potential has been fitted
to the $^{11}$Li r.m.s.\ radius and binding energy in {\em variational
calculations} that are unable to reproduce these quantities with
high accuracy.\\
\subsection{The soft dipole mode}
\par
The dipole transition operator in our model is of the form
\begin{eqnarray}
{\cal M}(E1\,\mu) = - \frac{N_{v}\,Z}{A}\;e\,y\,Y_{1\,\mu}(\hat{y})\,,
 \label{22}
\end{eqnarray}
where $e$ is the proton charge, $A=11$, $Z=3$ and the number of
valence neutrons, $N_{v}=2$.
The operator \mbox{(\ref{22})} corresponds to the
excitation of the three-body cluster  modes only. The
excitation energy of the first excited state of $^{9}$Li is
relatively high ($\sim$4 MeV). So, low-energy $E1$-transitions
correspond to the excitation of the cluster degrees of freedom
only and should be described by the operator \mbox{(\ref{22})}.
\begin{figure}[t]
\epsfverbosetrue
\epsfbox{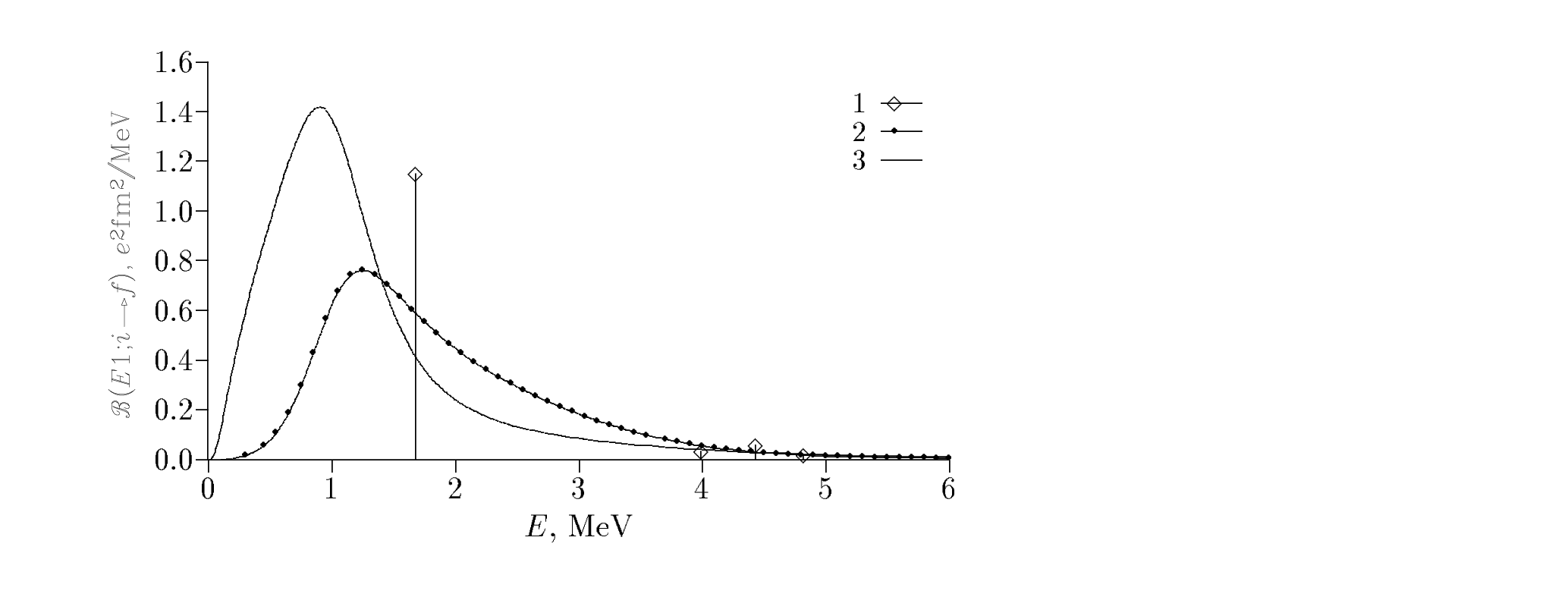}
\caption{Cluster ${\cal B}(E1; g.s. \to continuum)$ for
$^{11}$Li. Zero-width peaks with diamonds on the top (1)
given in arbitrary units have been obtained
in the pure variational approach for both ground and exited states, i.e.
the wave functions for the initial and the excited states have been
obtained by the diagonalization of the truncated Hamiltonian matrix.
The curve with dots (2) has been calculated with the variational ground
state wave function, the continuum spectrum wave function has been obtained
by the $J$-matrix approach. The solid curve (3) is the result of $J$-matrix
calculations for both the ground and continuum spectrum states.}
\end{figure}
\par
The cluster reduced probability of the $E1$-transition,
${\cal B}(E1;\;E_{f}-E_{0})$, associated with the operator (\ref{22}),
is displayed on the figure 2. In $J$-matrix approach, zero-width
pure variational peaks of ${\cal B}(E1;\;E_{f}-E_{0})$
gain finite widths and shift to lower energies.
It is seen that in calculation of ${\cal B}(E1;\;E_{f}-E_{0})$ it is
important to allow for the asymptotics not only for continuum states, but
for the ground state as well. Due to the wide spatial distribution of the
halo neutrons density,  large distances contribute
essentially to the $E1$ strength in the low-energy region. To account
for this contribution and to get a
convergence,  one should allow  for
ground and continuum state wave function
components with very large values of the
total number of oscillator quanta
$N \simeq 2000$\footnote{The
classical turning point for the basis function
with the total number of oscillator quanta $N=1000$ is at a distance of
$\approx$108 fm from the origin.}
in calculations of ${\cal B}(E1;\;E_{f}-E_{0})$ .
The large-distance $E1$-transitions
enhance the peak of the cluster ${\cal B}(E1;\;E_{f}-E_{0})$ and shift it
to a lower energy. Obviously, this peak should be associated with  the
soft dipole mode. Thus, the neutron halo manifests itself in the
appearance of the soft dipole mode that arises from the shift and
enhancement of the ${\cal B}(E1;\;E_{f}-E_{0})$ peak.
This effect appears to be very
important in the calculation of the $^{11}$Li electromagnetic dissociation
cross section (see below).
\par
Large distances contribute essentially to $E2$- and $E0$-transitions, too.
The shift and the enhancement of  ${\cal B}(E0;\;E_{f}-E_{0})$ and of
${\cal B}(E2;\;E_{f}-E_{0})$ are even more pronounced. Nevertheless,
$E2$- and $E0$-transitions do not play an important role in the
electromagnetic dissociation, and we shall not discuss them here.
\begin{figure}[tbh]
\epsfverbosetrue
\epsfbox{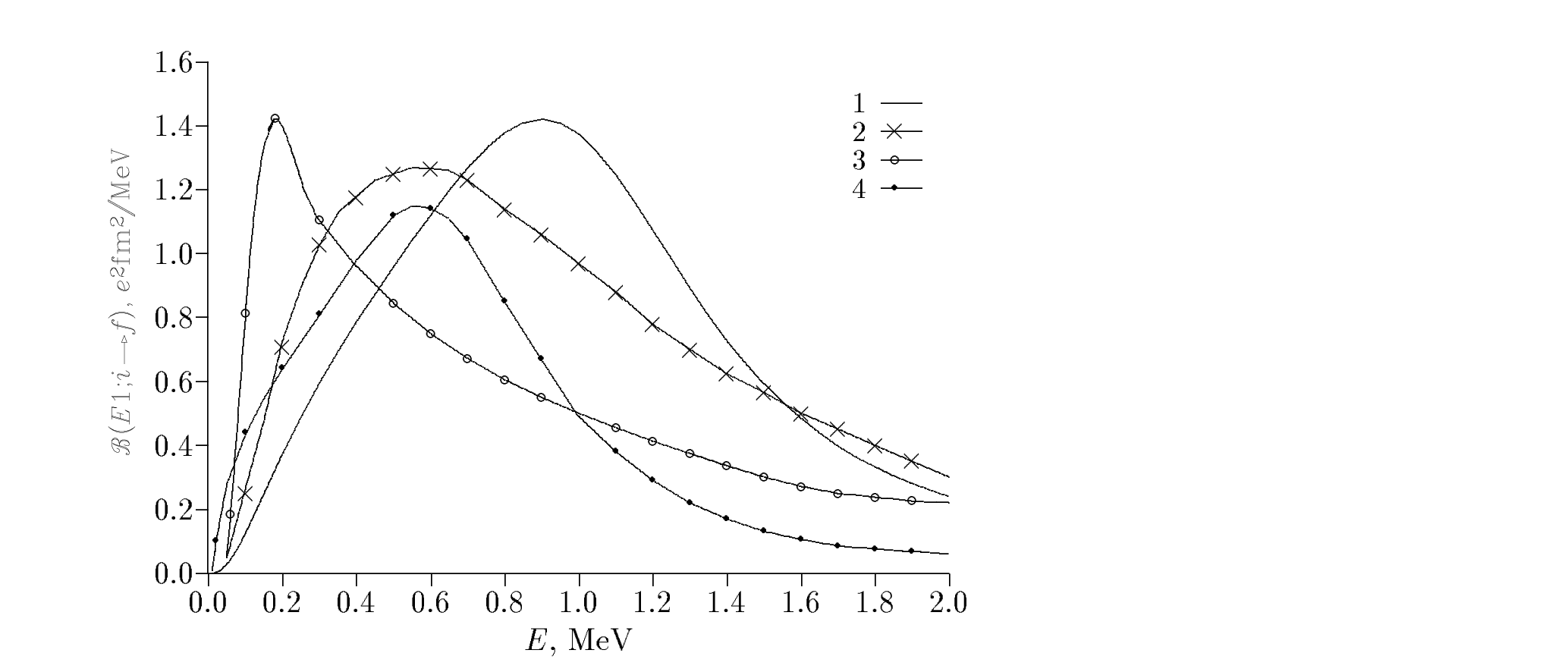}
\caption{
Comparison of our results for ${\cal B}(E1; g.s. \to
continuum)$ in $^{11}$Li with results of other authors. 1 --- this work
($J$-matrix method), 2 --- ref.\ \protect\cite{Danilin}, 3 ---
ref.\ \protect\cite{15}, 4 --- experimental data
parametrization of ref.\ \protect\cite{cc4}.}
\end{figure}
\par
Figure 3 shows the comparison of
the results of our calculations of cluster ${\cal B}(E1;\;E_{f}-E_{0})$
with the parametrization of experimental data of
ref.\ \cite{cc4}. The agreement is reasonable. The form of the
 ${\cal B}(E1;\;E_{f}-E_{0})$ peak is well reproduced, the discrepancy in
the position of the  ${\cal B}(E1;\;E_{f}-E_{0})$ maximum is supposed
to be eliminated by the adjustment of  the
potentials. The results of the  ${\cal B}(E1;\;E_{f}-E_{0})$ calculations of
refs.\ \cite{15,Danilin} are also depicted on fig.\ 3. The three-body
cluster calculations with the allowance for democratic decay channels of
ref.\  \cite{Danilin} nicely reproduce the energy of the soft dipole mode.
Nevertheless, the form of the peak in our calculations is reproduced
better. Note, that the authors of ref.\ \cite{Danilin} used
another set of the potential parameters. The calculations of
ref.\ \cite{15} with the allowance for two-body decay channels failed to
reproduce both the position and the form of the
 ${\cal B}(E1;\;E_{f}-E_{0})$ peak.
\par
The soft dipole mode exhausts about 90\%
of the cluster
EWSR associated with the operator \mbox{(\ref{22})},
$S_{\hbox{clust}}(E1)$. Nevertheless, using the results of
ref.\ \cite{EWSR} it is easy to obtain that
\begin{eqnarray}
\frac{S_{\hbox{clust}}(E1)}{S_{\hbox{tot}}(E1)} =
\frac{1}{12} \approx 8.3\%\, ,
\end{eqnarray}
where ${S_{\hbox{tot}}(E1)}$ is the total EWSR
accounting for excitations of all nucleons.
So, the contribution from the soft dipole mode to the total
EWSR is relatively small. In the vicinity of the sharp
${\cal B}(E1;\;E_{f}-E_{0})$ maximum
at the excitation energy $E\approx$1-2 MeV only $\sim$8\% fraction of the
total EWSR
is exhausted. Nevertheless,
the account for the soft dipole mode  results in an essential
increase
of the electromagnetic dissociation cross section
of 0.8 GeV/nucleon $^{11}$Li beams on Pb and Cu targets.
The wide space distribution of the halo neutrons density
is also
well-manifested in the electromagnetic dissociation
of $^{11}$Li beam.
The contribution of the large-distance $E\lambda$-transitions
to the cross section is about 50\%. The
calculations have been performed by the equivalent
photon method \cite{23}. The only parameter of the
method is the minimal value of the impact parameter $b_{min}$.
We use for $b_{min}$ the values of 9.0 fm for Pb and 6.8 fm for Cu target
nuclei, respectively.
These quantities are the sums of the $^{11}$Li and target nucleus charge
radii. With these values of $b_{min}$ we obtain for the
electromagnetic dissociation cross sections the
values of 0.966 barn for the Pb target and 0.132 barn for the Cu target; the
corresponding experimental values are $0.890\pm 0.110$ barn and
0.21$\pm$0.04 barn, respectively \cite{4}.
$E0$- and $E2$-transitions give only 1.2\%
contribution in the cross sections.
\par
$E1$-transitions in $^{11}$Li have been
studied  in the framework of RPA+two-body-con\-ti\-nu\-um model in
refs.\ \cite{11,lenske}.
Though our model assumptions differ significantly from the ones of
refs.\ \cite{11,lenske},
the results are in good agreement. For example,
the excitation energy values corresponding to the peaks of the function
$B(E1;\;E_{f}-E_{0})$ displayed on the fig. 2, are very close to the
values that one can find in refs.\ \cite{11,lenske}.
\section*{Conclusions}
\par
It is shown, that cluster model $^{9}$Li+$n$+$n$ yields a good
description of the ground state properties and $E1$-transitions in the
$^{11}$Li nucleus. The oscillator functions expansion technique may
be used in the studies of weakly-bound systems with long-tailed
wave functions, e.g., in the study of neutron halo properties.
For both bound and continuum states the correct
account of the wave function asymptotics in the framework of
the oscillator representation of scattering theory is very important
in such studies. Low-energy $E1$-transitions in $^{11}$Li are
of the cluster nature. The widths and the position
of resonant states calculated
in the democratic decay approximation are in a reasonable agreement with
experiment.
\vspace{3mm}
\par
\noindent
{\bf Acknowledgments}\\
\noindent
 We are thankful to Profs. J.Bang, B.Danilin,
I.Thompson and  J.Vaagen for valuable
discussions.

\end{document}